\documentclass[twocolumn,showpacs,preprintnumbers,amsmath,amssymb]{revtex4}

\usepackage{graphicx,color}
\usepackage{bm}

\begin{document}

\preprint{}
\bibliographystyle{unsrt}

\title{Microphase morphology in 2D fluids \\under lateral confinement}

\author{Alessandra Imperio$^1$
}
\author{Luciano Reatto$^2$}
 \affiliation{$^1$ CNISM, Sezione dell'Universit\`a degli Studi di Milano, Italy.\\
$^2$ Dipartimento di Fisica, Universit\`a degli Studi di Milano,via Celoria 16, 20133 Milano, Italy
}


\date{\today}

\begin{abstract}
We study the effects of confinement between two parallel walls on a
 two dimensional fluid with  competing interactions which lead to the formation of particle micro-domains at the thermodynamic equilibrium (microphases or microseparation). The possibility to induce structural changes of the morphology of the micro-domains is explored, under different confinement conditions and temperatures. In presence of neutral walls, a switch from stripes of particles to circular clusters (droplets) occurs as the temperature decreases, which does not happen in bulk. While the passage from droplets to stripes, as the density increases, is a well known phenomenon, the change of the stripes into droplets as an effect of temperature is rather unexpected. Depending on the wall separation and on the wall-fluid interaction parameters, the stripes can switch from parallel to perpendicular to the walls and also a mixed morphology can be stable.

\end{abstract}

\pacs{Valid PACS appear here}
\maketitle

The formation of spatially modulated phases at the thermodynamic equilibrium (microphases or microseparation) is a phenomenon occurring in very different systems. Examples of 2D microphases can be found in adsorption films on a solid substrate \cite{plass02,kern91} and in many dipolar fluid films \cite{klokkenburg06,hoffmann06,schneider05,brown01,seul95}. Examples in 3D involve colloidal suspensions \cite{lu06, stradner04} and amphiphilic systems such as block copolymers \cite{bates99,fink99}. Typical 2D particle domains include droplets, stripes, strings, rings, bubbles, forming either regular or disordered configurations. In 3D, instead, spherical clusters, array of cylinders and lamellae are obtained. Interest in such phenomena is very high, also for the number of possible applications in nanotechnology and material science, where new devices with specific electric, optical and rheological properties are designed. For example, as the 2D patterns can be easily transferred on different substrates, they can be used as masks in nanolitography \cite{choi99}. Moreover, connections between the general mechanisms governing the microphases and other branches of science are currently under investigation: from vitrification and gelation \cite{charbonneau06,decandia06,sciortino03}, to particle aggregation in colloidal solutions \cite{lu06,stradner04,campbell04}, to the aggregation of proteins attached to biological membranes \cite{destainville06} and to the cluster formation in various disease processes.\\\indent The presence of anisotropic forces often leads to the formation of micro-domains. In the last few years it has been recognized that also an isotropic interaction can lead to microphases. For instance, in colloidal systems, which this work is addressed to, the competing interactions scenario is often invoked to explain the formation of micro-domains. This model is based on the use of an effective potential, describing the interparticle interactions mediated by the environment, which can be a solvent, a substrate or an external field. Within this frame, the effective potential is made up of a short-range attraction (for instance, due to depletion forces) plus a longer-range repulsion (for instance, due to partially screened electrostatic forces): the attraction favouring the condensation of particles, while the longer-range repulsion limiting the cluster growth, as firstly pointed out in \cite{lebowitz65}. The cluster morphology depends on the ranges of the competing interactions and on the density, but also the temperature can have a role. It is interesting to notice that other forms of the interparticle potential lead to cluster formation, such as the softened-core models in \cite{malescio03} and the generalized repulsive interactions in \cite{mladek06}.\\
\indent While the theory of simple fluids under confinement is well developed \cite{evans90,gelb99}, the general behaviour of complex fluids, such as those undergoing microseparation, has received very limited attention. It is all the same clear that confinement is expected to be a powerful tool to modify and even induce new pattern morphologies, which might not appear at all in the bulk. This has been shown for some studies of block copolymers between smooth or patterned substrates \cite{yu06,tsori06,li06,duchs04,tsori01}. We have no knowledge, instead, of studies related to the confinement of colloidal systems developing microphases, which is the situation under investigation in this paper. In particular we treat the 2D case, so that this study is relevant, for instance, for Langmuir monolayers: an insoluble film of particles, entrapped at the interface of two fluids such as air and water. Moreover, the use of lateral barriers is a typical tool to influence the system, such as modifying the surface particle concentration untill the collapse of the monolayer.\\
\indent In our model the particle-particle potential has an impenetrable core of diameter $\sigma$, followed by an attractive well and a longer-range repulsive tail. The specific form we use reads:
\vspace{-0.4cm}
\begin{equation}
U_{pp}(r)= \left\{\begin{array}{ll}
                               \infty& \mbox{if $r< \sigma$} \\
                                U(r)-U(R_{cut}) & \mbox{if $\sigma\leq r\leq R_{cut}$}\\
                                   0 & \mbox{otherwise}\\
       \end{array}
\right.
\label{Uff_1}
\end{equation}
\vspace{-0.6cm}
\begin{equation}
U(r)=-\frac{\epsilon_a\sigma^2}{R_a^2}\exp(-\frac{r}{R_a}) +\frac{\epsilon_r\sigma^2}{R_r^2}\exp(-\frac{r}{R_r}),
\label{Uff_2}
\end{equation}

\noindent  $r$ being the interparticle distance, $R_{cut}$ the separation at which the potential is cut and shifted to zero. The potential parameters are: $R_a~=~1~\sigma,\epsilon_a~=~1$ for the short-range attraction and $ R_r~=~2~\sigma, \epsilon_r~=~1$ for the longer-range repulsion; the cutoff is $R_{cut}~=~10$. Apart for a constant term in eq.~(\ref{Uff_1}), $U_{pp}(r)$ is the same interaction studied at first in \cite{sear98} and then in \cite{imperio04,imperio06,destainville06} for bulk 2D fluids. The fluid is confined by smooth parallel walls, set at $x~=~\pm ~L_x/2$, $L_x$ being the length of the simulation box side. Periodic boundary conditions are implemented in the $y$ direction. As wall-particle interaction we assume the following form:
\begin{equation}
U_{wp}(x)=\left\{\begin{array}{ll}
                               \infty& \mbox{if $|x|>\frac{L_x-\sigma}{2}$} \\
			       V(x)-V(x=0) & \mbox{otherwise}\\
                \end{array}\label{Uwf_1}
\right.
\end{equation}
\footnotesize
\begin{equation}
\hspace{-0.5cm}V(x)= \alpha U_0 \left[\exp\left(-\frac{L/2-\sigma/2}{R_{wp}}\right)\left(\exp\left(-\frac{x}{R_{wp}}\right)+\exp\left(\frac{x}{R_{wp}}\right) \right)\right]
\label{Uwf_2}
\end{equation}
\normalsize
 The quantity  $\alpha$ is a dimensionless factor, $U_0~=~|U_{pp}(r~=~\sigma)|$, so that the wall-particle potential energy for a particle in contact with the wall is $V_0~=~V~(x=L_x/2-\sigma/2)\sim \alpha ~U_0$. $R_{wp}$ is the range of the wall-particle potential. When $\alpha=0$ we have two neutral hard walls at distance $L_x$; $\alpha~>~0$ represents repulsive walls. Since we want to study the generic behaviour of the system, it is convenient to use as control parameter the temperature $T$. In specific cases, the control parameter has a different meaning, like the concentration of non adsorbing polymers with the depletion interaction. Hereafter, every physical quantity is expressed in reduced units: energies and temperatures in units of $U_0$, lengths in units of $\sigma$, densities in units of $\sigma^2$, specific heat in units of $k_B$ (the Boltzmann constant). Montecarlo simulations are performed in the $NVT$  ensemble ($N$ number of particles, $V$ volume, $T$ temperature). Specific heat data are obtained through the Parallel Tempering (PT) \cite{bibbia} technique.\\ \indent Microphase pattern depends on the particle mean density $\rho$: in bulk, the system goes from droplets to stripes \cite{imperio06} to bubbles as $\rho$ increases. In this letter we focus our attention mainly on the striped phase. For example, at $\rho~=~0.4$, the bulk fluid with potential (\ref{Uff_1}-\ref{Uff_2}) develops a grid of parallel stripes, the period of which is $P~ =10.8~\pm ~1.5$. In a previous study \cite{imperio04} we have shown that the value of the cutoff $R_{cut}$ has a negligible effect on the period, for instance, an increase of 50$\%$ in $R_{cut}$ changes $P$ by less than the uncertainty on the value of P.  The transition, from the homogeneous phase to the striped phase, is signaled by a peak in the specific heat, around the temperature $T^*\approx~0.58$. At lower $T$, around 0.4, the particles order on a triangular lattice within the stripes. We will show that the confined system behaves differently for $T~<~\theta$ or $T~>~\theta$ with $\theta \approx 0.4$.\\
\indent The behaviour of the system at $\rho=$ 0.4, with neutral walls ($\alpha~=~0$), is firstly analyzed for intermediate temperatures ($\theta<T<T^*$).
\begin{figure}
\includegraphics[width=8.6cm]{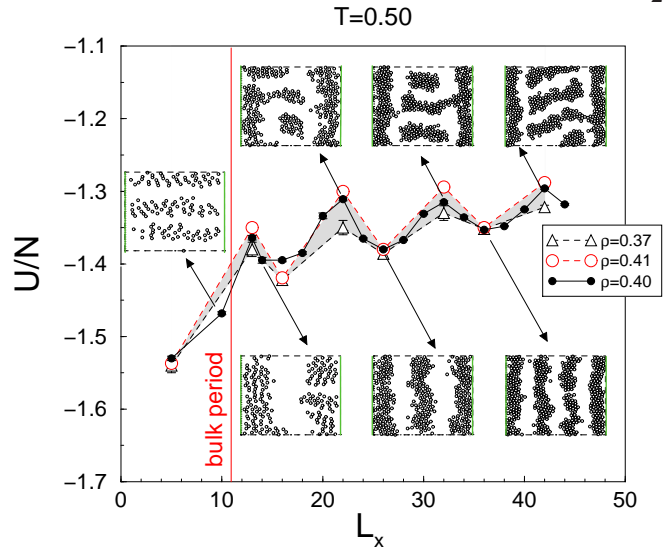}
\caption{\label{U_Lx}\footnotesize{Potential energy vs wall separation for three densities at $T=0.5$. The system is confined between two neutral hard walls along the $x$ direction. The insets show some snapshots at $\rho=0.4$. The vertical line represents the stripe bulk period $P$. The shaded area emphasizes the oscillating profile of $U$ for $\rho=0.37$ and 0.41.} }
\end{figure}
The general trend of the potential energy is to increase with $L_x$ (Fig.~\ref{U_Lx}) towards the bulk value. This is related to the growth of the fraction of particles (those in the centre) which do not ``see'' the walls, so that they behave as in the bulk. On the contrary, the particles nearby the walls feel a reduced  longer-range repulsion, since they interact with fewer particles: their energy is therefore more negative than in bulk. What is remarkable is that the energy profile at fixed $T$ exhibits a number of maxima and minima as function of $L_x$. These extrema of $U$ are connected to different patterns (insets in Fig.~\ref{U_Lx}) stabilized by the walls. For $L_x\leq ~12$, the stripes alignment is perpendicular to the walls. We stress that, even if $L_x$ is large enough to fit in one parallel stripe, that never happens: the system switches abruptly from perpendicular stripes to two parallel stripes. 
For $L_x~>~15$: energy maxima correspond to the most frustrated configurations, where two stripes form close to the walls and perpendicular ones in the middle; energy minima correspond to the unfrustrated parallel stripes, occurring for wall separations $L_x~\approx ~n P + \Delta$ ($n$ integer, $P$ bulk period, $\Delta$ stripe width). Similar results are found also for slightly different densities such as $\rho=0.37-0.41$ and plotted in Fig.~\ref{U_Lx}.\\
\begin{figure}[ht!]
\includegraphics[width=8.6cm]{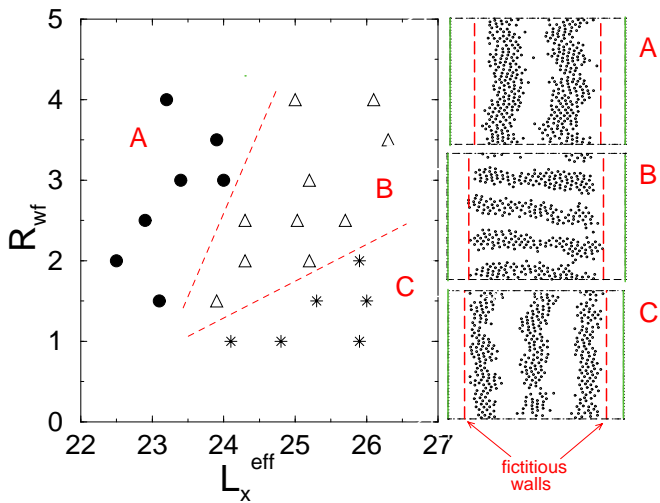}%
\caption{\label{diagramma1}\footnotesize{Stripes morphology: ($\bullet$) two stripes, ($\triangle$)~stripes perpendicular to the walls, ($*$)~three~stripes, as a function of the wall-particle potential range $R_{wp}$ and of the effective wall separation $L_x^{eff}$. All simulations are performed with the nominal density $\rho=0.3$ and $T=0.5$, using $N=400$ and $L_x=32$. The effective densities, $\rho^{eff}$, range from 0.37 to 0.41.
} }
\end{figure} 
\indent Switching on the wall-particle repulsion ($\alpha~>~0$), we find a striking dependence of the stripe morphology on the parameters of $U_{wp}(x)$. In this case the particles are squeezed in the central region, so that the space truly accessible is reduced (right panels in Fig.~\ref{diagramma1}),  and the system is shifted at an effective density greater than the nominal one. We might think to replace the repulsive walls by two fictitious neutral walls at distance $L_x^{eff}$.  We estimate $L_x^{eff}$ and $\rho^{eff}$ in the following manner: for each configuration picked up for averages, we search the smallest and highest value of $x$ and put $L_{x}^{eff}=x_{max}-x_{min}+\sigma$, and then $\rho^{eff}=N/(L_x^{eff}L_y)$. A fixed value of $L_x^{eff}$ can be obtained with different combinations of the parameters $\alpha$ and $R_{wp}$ in $U_{wp}(x)$. If $R_{wp}(x)\lesssim L_x/4$, $L_x^{eff}$ can be estimated as the length at which the following relation holds:
\begin{equation}
U_{wp}(|x|=L_x^{eff}/2) + U^*= T,
\label{Lx_eff_t}
\end{equation}
 $U^*$ being the bulk energy per particle of $\rho^{eff}$. The discrepancy between the values of $L_x^{eff}$ obtained through  eq.~\ref{Lx_eff_t} and via simulations is less than $5-6\%$ for the cases shown in Fig.~\ref{diagramma1}.\\
\indent Now, let us discuss in parallel the patterns obtained with true neutral walls as a function of $L_x$ (Fig.~\ref{U_Lx}), and those corresponding to fictitious neutral walls with $L_x^{eff}=L_x$ (Fig.~\ref{diagramma1}). First of all, with $L_x=26$ the potential energy profile has a minimum corresponding to the formation of three stable parallel stripes; with $L_x^{eff}=26$, instead, we can stabilize parallel or perpendicular stripes depending on $R_{wp}$ (Fig.~\ref{diagramma1}). Interestingly, such a change occurs for $R_{wp}\sim R_r$, i.e. the interparticle repulsion range. As $L_x$ decreases towards 22, the system with true neutral walls moves towards a frustrated configuration having both parallel and perpendicular stripes. As $L_x^{eff}$ decreases towards 24, the stripes switch from parallel to perpendicular for smaller values of $R_{wp}$; finally at $L_x^{eff}\sim 23$ the parallel orientation is once again preferred, but it shows significant changes in the stripes width and period, with respect to the bulk and even to the case with neutral walls. In particular, in the region C of Fig.~\ref{diagramma1} $\Delta$ is not affected by the repulsion while $P$ can vary up to 20$\%$ with respect to the bulk; in the region B, $P$ is not modified while minor changes of $\Delta$ up to $8\%$ are present; in the region C both $P$ and $\Delta$ are modified up to 25-30$\%$.\\ 
 We conclude that strong confinement not only affects the orientation of the stripes but also the basic parameters, period $P$ and $\Delta$, of the microphase. In this respect the behaviour of our system is rather different from the one reported in \cite{yu06}, in which asymmetric block copolymers are confined in a cylindrical pore of diameter $D$. When cylinders form within the neutral pore, those are parallel or perpendicular to the wall depending whether $D$ is a multiple of the bulk period or not. When the confining pore wall becomes attractive or repulsive a wide number of different morphologies is obtained in \cite{yu06}, the period of which is the same as in bulk. In our case, instead, the morphology at intermediate $T$ remains essentially striped, but the values of $P$ and $\Delta$ show some variation depending on the particle-wall interaction. A possible origin of such different behaviour in the two systems under confinement is that in case of ref. \cite{yu06} the diameter of the cylinders is much smaller of the period of the array whereas in our system $\Delta$ and $P$ are comparable 

\begin{figure}
\includegraphics[width=8.6cm]{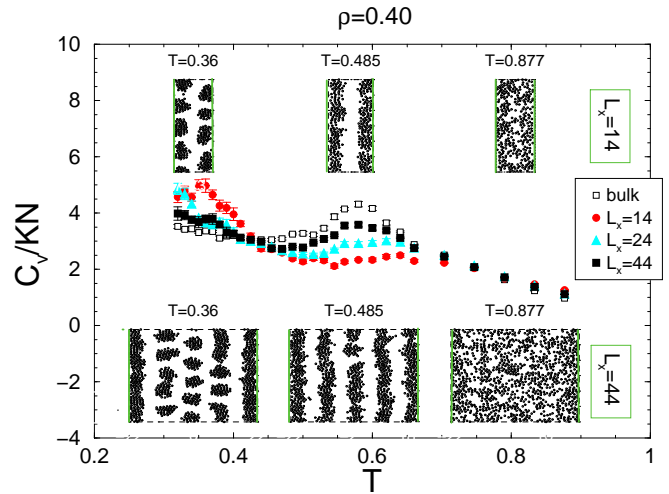}
\caption{\label{Cv2} \footnotesize{(color online) Specific heat at constant volume per particle vs temperature for bulk and for different wall separations $L_x$. $C_v$ is obtained as energy derivative. Confined systems are subject to neutral hard walls. Snapshots of the fluid at different temperatures are plotted in the insets. Periodic boundary conditions are applied along the $y$ direction (dashed lines in the insets).} }
\end{figure}
We find that also the temperature has a noticeable effect on the microphase morphology under confinement. In bulk the onset of microphases is marked by a peak in the specific heat. In systems confined between neutral walls, such a peak becomes broader and reduced in amplitude, untill for $L_x=14$ it is barely a plateau. At the same time a secondary peak develops at low temperatures ($T\sim 0.36$), for very narrow systems ($L_x=14$), with a droplet-stripe passage driven by temperature (snapshots at low $T$ in Fig.~\ref{Cv2}). The results for $L_x=24$ suggest that such secondary peak in the specific heat is still present but shifted at a temperature below the lowest $T$ of our computations. Likely such a peak disappears when $L_x$ is sufficiently large, since the system resembles more and more the bulk one, where the stripes are the most stable phase, as discussed in \cite{imperio04}. Increasing the wall separation, the droplets persist at low $T$ mainly in the central region, while two stripes form close to the walls. The droplets are ordered as it is proved by the presence of Bragg peaks at short wave vectors in the static structure factor. Circular clusters are entirely replaced by stripes for $T\gtrsim 0.4$.\\
In the bulk, the period slightly changes with the temperature \cite{imperio06}. Therefore one might suspect that the droplet formation is  due to a change of $P$ with $T$, so that a given $L_x$ is commensurate to $P$ at a certain $T$ while it is not at another one. This is not the case as shown by low $T$ simulations with $L_x$ ranging from 13 to 19:  droplets always form at low $T$. We are therefore confident that such droplet formation at low $T$ is not due to frustration like that governing the stripe orientation between neutral walls. In conclusion, the droplets can occur under two conditions: i) at fixed low $T$, droplets are favoured by strong confinement; ii) at fixed $L_x$, droplets appear at low enough temperatures.
The role of energy and entropy in the stripe orientation as well as in the droplet formation will be discussed in a more extensive paper.\\
\indent In summary we have analysed a 2D microseparated fluid under lateral confinement, at a density such that stripes are formed in the bulk. We find that, with neutral walls, stripes perpendicular to the walls are favoured for $L_x\lesssim P+\Delta$, $P$ bulk period and $\Delta$ stripe width. Unfrustrated parallel stripes are obtained for wall separation $L_x\approx nP+\Delta$ ($n$ integer). Otherwise frustrated configurations, made up of parallel stripes close to the walls and perpendicular stripes in the central region are stabilized, in contrast with the bulk case. The effect of a repulsive wall cannot be represented by a simple reduction of the space $L_x^{eff}$ available to the particles. In fact for a given $L_x^{eff}$, different patterns can be stabilized depending on the wall-particle potential range. This is not possible with neutral walls, since a unique configuration is selected for each separation $L_x$. Moreover, in presence of repulsive walls, a striped phase can be stabilized, characterized by rather different values of $P$ and $\Delta$ with respect to the bulk ones. Finally, the role of temperature is important and  droplets are favoured at low $T$ for a system under confinement.\\  Our results show that strong confinement is a way to control the microphase morphology also in unexpected ways.  The easiness in switching the stripe orientation at fixed $T$, by tuning the wall-particle interaction, might be an interesting property to investigate in materials which are also optically active. Finally it is of general interest to test the stability of the pattern morphology on the sample temperature as well as on the degree of confinement, in order to develop new materials.\\
\indent\textsl{ This work is supported by the MCRT Network on Dynamical Arrest. The authors thank dr. A. Corsi for useful discussions.}

\end{document}